\documentstyle[twoside,fleqn,espcrc2]{article}
\def\p{\partial} \def\D{{\cal D}}
\def\dpp{\p^{++} }  \def\Dpp{\D^{++} }
\def\e{\epsilon} \def\g{\gamma} \def\n{\nabla} \def\d{\delta}
    
  \def\bt{{\bar \vartheta}} 
\def\b{\beta} \def\a{\alpha} \def\l{\lambda} \def\m{\mu} \def\f{\varphi}
\def\G{\Gamma} \def\L{{\cal L}^{+4} }
\def\da{{\dot \alpha}} \def\db{{\dot \beta}}  \def\dg{{\dot \gamma}}
\def\dd{{\dot \delta}} \def\dm{{\dot \mu}}

\def\o{\omega} 
\def\half{{1\over 2}} \def\N#1{$N\!=\!#1$ }
\def\der#1{{\partial \over \partial #1}}
\def\pp#1#2{{\partial #1 \over \partial #2}}
\def\r#1{(\ref{#1})}
\font\sqi=cmssq8
\def\C{\kern2pt {\hbox{\sqi I}}\kern-4.2pt\rm C}
\begin{document}
\title{
\begin{flushright}hep-th/9602058\end{flushright}
Unravelling the On-shell Constraints of Self-dual Supergravity Theories}
\author{Ch. Devchand\thanks{
Talk presented by Ch. Devchand at the 29th International Ahrenshoop
Symposium on the Theory of Elementary Particles,
Buckow, August 1995}
and  V.  Ogievetsky \\
Joint Institute for Nuclear Research, 141980 Dubna, Russia}

\begin{abstract}
We review a construction, using the harmonic space method, of solutions
to the superfield equations of motion for N-extended self-dual
supergravity theories. A superspace gauge condition suitable for the
performance of a component analysis is discussed.
\end{abstract}
\maketitle

There has recently been some interest in self-dual supergravity theories
in the context of string theory. Specifically, in the context of the string
with local \N2  world-sheet supersymmetry, which was already recognised by
its discoverers \cite{a} as having a finite spectrum and therefore just a
fancy way of writing down a field theory. Although this string theory,
with critical dimension $2+2$, has been around since the mid-70's, it was
only about five years ago that Ooguri and Vafa \cite{ov} appreciated that
they contained Yang-Mills and gravity and not just scalars. Siegel \cite{s}
then suggested that the field theories which closed \N2 strings
(and their heterotic versions) describe are actually supersymmetrisations
of the self-duality conditions for the Riemann tensor. This talk is however
not about these string theories (recent progress was described at this
symposium by O. Lechtenfeld), but about the equations of motion of self-dual
supergravity theories, which are interesting in their own right \cite{sdsg}.
In particular, these equations are a very convenient way of describing
hyperk\"ahler spaces with (extended) Poincar\'e supersymmetry. They are
also a further large class of 4 dimensional models which are to some extent
exactly soluble.

To establish notation, let us start with the standard \N0
(non-supersymmetric) self-duality equations \cite{sdg,gios}.
In 2-spinor notation, which we shall use, the covariant derivative with
respect to a vierbein takes the form
\begin{equation} \n_{\alpha \da} =
E^{\mu\dm}_{\a\da}\partial_{\mu \dm}
+ \omega_{\alpha\da}\   ,\label{1}\end{equation}
where $\mu , \dm$ are world and $\a ,\da$ are tangent-space indices. The
connection $\omega_{\alpha\da}$ takes values in the algebra of tangent-space
rotations. For generality this can be taken to be $sl(2,C)\times sl(2,C)$.
Appropriate hemiticity conditions for either the Euclidean (4,0) or Kleinian
(2,2) signatures can be imposed. In the absence of torsion,
\begin{equation}
\n_{\alpha \da}E^{\mu\dm}_{\b\db} - \n_{\b\db}E^{\mu\dm}_{\alpha \da} = 0
\label{tor}\end{equation}
and the components of the connection $\o_{\a\da}$ are completely determined
in terms of the vierbein and its inverse $E_{\mu\dm}^{\alpha \da};\quad
E^{\mu\dm}_{\alpha \da}E_{\mu\dm}^{\beta \db} = \d^\b_\a \d^\db_\da$, e.g.
the components in the dotted $sl(2)$ are
\begin{equation} (\o_{\a\da} )_\db^\dd =
    - \e_{\a\b} E^{\b\dd}_{\m\dm}\p^\g_{(\da}E^{\m\dm}_{\g\db)}
 - {1\over 3} \e_{\da\db} E^{\b\dd}_{\m\dm}\p^\dg_{(\a}E^{\m\dm}_{\b)\dg}\
  .\label{2}\end{equation}
The components of the Riemann tensor are given by the commutator
\begin{equation} [\D_{\a\da} ~,~\D_{\b\db}] =
        \e_{\da\db} R_{\a\b} + \e_{\a\b} R_{\da\db}\   ,
\nonumber\end{equation}
where the decomposition in the basis $\{ \G^{\a\b} ,\G^{\da\db}\}$ of the
tangent space algebra reveals the irreducible components of the Riemann
tensor
\begin{eqnarray}
R_{\alpha \beta} &\equiv &
C_{(\alpha\beta\gamma\delta)} \Gamma^{\gamma\delta}
 + R_{(\alpha\beta)(\dg\dd)} \Gamma^{\dg\dd} +
{1\over 6}R \Gamma_{\alpha\beta} ,\nonumber\\[5pt]
R_{\da \db} &\equiv &
C_{(\da\db\dg\dd)}\Gamma^{\dg\dd} + R_{(\g\d)(\da\db)} \Gamma^{\gamma\delta}
+ {1\over 6} R \Gamma_{\da\db},\nonumber\end{eqnarray}
where $C_{(\da\db\dg\dd)} (C_{(\alpha\beta\gamma\delta)}) $ are the
(anti-) self-dual parts of the Weyl tensor, $R_{(\alpha\beta) (\dg \dd)}$
are the components of the tracefree Ricci tensor, $R$ is the scalar
curvature. In this notation self-duality of the Riemann
tensor takes the form
\begin{equation}  R_{\da\db} =0\  ,\label{sd}\end{equation}
which may equally be written in the fashion of the commutation relation
\begin{equation} [\D_{\a\da} ~,~\D_{\b\db}] = \e_{\da\db} R_{\a\b}  \  ,
\label{sdg}\end{equation}
where $R_{\a\b} = C_{\a\b\g\d}\G^{\g\d}$, since the above decomposition
clearly shows that \r{sd} is tantamount to the
vanishing of the Ricci tensor and the self-duality of the Weyl tensor.
Since the curvature takes values in only one of the two $sl(2)$ factors
of the tangent space algebra, the connection may also be chosen to have
the basis $\{\G^{\g\d}\}$; the components in the dotted
$sl(2)$, i.e. \r{2}, being manifestly gauge-artefacts.
In fact the condition
\begin{equation}
 (\o_{\a\da} )^\db_\dg = 0   \label{sdeich}\end{equation}
is a first-order equation for the vierbein which implies that the connection has
self-dual curvature. Modulo gauge transformations, it is clearly
equivalent to the condition \r{sd}\cite{eh}. This makes manifest the
equivalence between the hyperk\"ahler restriction of the holonomy group
to $sl(2)$ and self-duality of the Riemann tensor. In the self-dual gauge
\r{sdeich} the conditions \r{sdg} are therefore not dynamical equations
for the connection. The dynamics is in fact described entirely by the
differential part of \r{sd}, i.e. by the zero torsion constraint for the
vierbein \r{tor}. Self-duality is guaranteed by the fact that the tangent
algebra (now just $sl(2)$) does not act on dotted indices, the second $sl(2)$
(with generators $\{\G^{\dg\dd}\}$) now acting only globally. There is now no
distinction between dotted world and tangent-space indices.
Once \r{sdg} has been solved for the vierbein $E^{\mu\dm}_{\a\da}$,
the self-dual metric is obtained immediately from its inverse
\begin{equation} ds^2 =
\e_{\a\b}\e_{\db\dd}E^{\a\db}_{\m\da}E^{\b\dd}_{\nu\dg}dx^{\m\da}dx^{\nu\dg}.
\end{equation}

The remarkable thing about self-dual theories is that supersymmetrisation is
{\it not} dependent on the number of independent supersymmetries. Equations
\r{sdg} can be supersymmetrised in a uniform fashion for any extension N,
merely by replacing the undotted $sl(2)$ indices by $osp(N|2)$ indices
corresponding to the local tangent-space of chiral superspace with
coordinates $z^{M \da}= \{ x^{\mu \da}, \bt^{m \da} \}$. Thus the
supertorsion constraints,
\begin{equation}  [\n_{B \db}, \n_{A \da}\} = \e_{\da \db} R_{AB}
                    \label{ssd}\end{equation}
encapsulate the dynamical equations for N-extended self-dual supergravity
\cite{s}. The chiral superspace covariant derivative is given in terms of
a supervierbein and a superconnection depending on the chiral superspace
coordinates $z^{M\da}$,
\begin{equation} \n_{A \da} = E^{M \db}_{A \da} \p_{M \db}
             + (\o_{A \da} )_{BC} \G^{BC} \  ,\label{cov}\end{equation}
where  $\G^{BC}$ are $osp(N|2)$ generators and
$\p_{M\db} = \{ \p_{\m\db}, \p_{m\db} = \der{\bt^{m\db}}\}.$
Note that unlike ordinary supergravities, for self-dual supergravities a
{\it supergroup} can be consistently gauged \cite{s}.
It is almost incredible that this remarkably simple generalisation of
\r{sdg} actually yields consistent equations. One ramification is that the
standard twistor transform \cite{p} for self-dual gravity may also be
generalised in a simple fashion to these super-extensions. An analytically
convenient variant of this transform uses harmonic spaces \cite{h}. It can
be described in an  N-independent manner, holding therefore for the \N0
case too. Our toolbox for solving the differential equations \r{ssd} for the
supervierbein consists of harmonics, i.e. a pair of constant spinors
$u^{+\da} , u^{-\da}$ satisfying the pairing condition
$\e_{\da\db}u^{+\da}u^{-\db} = 1$ and the equivalence relation
$u^{\pm\da} = e^{\pm\G} u^{\pm\da},\quad \G\in \C $, which basically says
that the $\pm$ indices on the $u$'s denote a $U(1)$ charge.
In this auxiliary space with coordinates $u^{\pm}$, the differential
operators
$$ \p^{\pm\pm} = u^{\pm\da} \der{u^{\mp\da}},\quad
   \p^0 = u^{+\da} \der{u^{+\da}} - u^{-\da} \der{u^{-\da}}  $$
realise the $sl(2)$ algebra.

Using these harmonics, we can define linear combinations of the covariant
derivatives \r{cov}
\begin{equation} \n^\pm_A  = u^{\pm\da} \n_{A \da} =
      u^{\pm\da} E^{M \db}_{A \da} \der{z^{M \db}} + \o^\pm_A\  ,
      \end{equation}
in terms of which the self-duality equations \r{ssd} are equivalent to the
following system of equations in the total space with independent variables
$\{z^{\pm M}, u^{\pm\db} \}$:
\begin{equation}\begin{array}{rll}
 [\n^+_A, \n^+_B] &=&0 \\[5pt]
 [{\cal D}^{++}, \n^+_A] &=& 0 \\[5pt]
 [\n^+_A, \n^-_B] &=& 0
 \quad\hbox{(modulo  $R_{AB}$)} \\[5pt]
[{\cal D}^{++}, \n^-_A] &=& \n^+_A ,\label{cr}\end{array}\end{equation}
together with consequences. The equivalence \r{ssd} $\Leftrightarrow$ \r{cr}
can be proven in the `central' coordinate basis
$z^{M\pm} \rightarrow z_c^{M\pm} = z^{M\da} u^\pm_\da $, in which
$\Dpp \rightarrow \dpp $.  The proof uses the fact that the equation
$\p^{++}l^+=0$ implies linearity of $l^+$ in $u^+$; all connections in
\r{cr} being defined globally on the harmonic two-sphere.
Clearly, the harmonics are useful in making flat
super\-planes manifest. Now, these are covariant equations in the enlarged
total space of $z$ and $u$ variables, and this system has enhanced
invariances: $u$-dependent transformations now being allowed. In particular,
since the $\D^+_A$'s commute amongst themselves, by Frobenius' theorem, we
may perform a {\it $u$-dependent} coordinate transformation
$z^{M\da}\rightarrow z_h^{M\pm}(z^{M\da}u^\pm_\da , u^{\pm\da})$
to a coordinate system in which $ \D^+_A \rightarrow \der{z^{A-}_h} $,
the flat partial derivative. Such a transformation also requires
a tangent transformation to a frame which preserves this flatness,
i.e. a frame in which the connection $\o^+_A$ is gauged to zero,
$\o^+_A \rightarrow \f^{-1}(\o^+_A + \p^+_A)\f = 0 $.
In such a Frobenius coordinate system the first equation in \r{cr} becomes
trivial and there is no distinction between world and tangent indices.
However, $\dpp \rightarrow\Dpp$, no longer a flat derivative,
\begin{eqnarray}
\Dpp &=& \dpp + (\dpp z_h^{M+})\der{z_h^{M+}} \nonumber\cr &~&+
(\dpp z_h^{M-})\der{z_h^{M-}} + \f^{-1}\dpp\f\  \nonumber\end{eqnarray}
and the problem actually reduces to solving the remaining equations in \r{cr}
for the vielbeins and connection of $\Dpp$,
\begin{equation} \begin{array}{rll}
          H^{++M+}  &:=&  \dpp z_h^{M+} \\[5pt]
	  H^{++M-}  &:=&  \dpp z_h^{M-} - z_h^{M-} \\[5pt]
           \o^{++}  &:=&  \f^{-1}\dpp\f\  .\label{H}\end{array}\end{equation}
These remaining equations, however, are completely soluble in terms of
an arbitrary {\it holomorphic prepotential}, the following expressions
providing their general solution:
\begin{equation} \begin{array}{rll}
	  H^{++M+}  &=&  \e^{MN} \p^-_N \L  \\[5pt]
	  H^{++M-}  &=&  \e^{MN} z^P_h \p^-_P  \p^-_N \L  \\[5pt]
    (\o^{++} )^M_P  &=&  \e^{MN} \p^-_P  \p^-_N \L        .\label{soln}
\end{array}\end{equation}
Here $\e^{MN} = (\e^{\mu\nu}, \eta^{mn})$ is the
graded\--skew\-symmetric $osp(N|2)$ invariant  and
$\L = \L(z^+_h,  u^\pm)$ is an arbitrary holomorphic (independent
of the $z^-_h$ coordinates) prepotential. In fact, in every gauge-equivalence
class of $\L$'s there is always one which has no explicit $u^+$-dependence,
so we may think of $\L(z^+_h,  u^-)$ as a {\it parametrisation} of self-dual
metrics: To every such $\L$ corresponds a self-dual metric. This of course
says nothing about regularity of the solution or boundary conditions (such as
asymptotic flatness, which is often of interest). These are separate
questions. In the \N0 ($\bt$-independent) case, $\L$ is also the appropriate
potential in the supersymmetric action for \N2 $D\!=\!4$ $\sigma$-models
having the corresponding hyperk\"ahler target manifold \cite{hsig}.

Having solved the equations \r{ssd} in  one coordinate system in this larger
space, we of course need to return to the original (`central') coordinates of
flat $\dpp$ derivatives in order to explicitly construct the metric
corresponding to any given $\L$. So we need to perform the reverse
coordinate transformation, $\Dpp\rightarrow \dpp$.
To do this, comparing the solution \r{soln} with the definitions \r{H},
we clearly need to solve the following system of first-order equations for
any specified choice of $\L$:
\begin{equation} \begin{array}{rll}
    \dpp z_h^{M+}  &=&  \e^{MN} \p^-_N \L                           \\[5pt]
 \dpp z_h^{M-}  &=&  z_h^{M+} + \e^{MN} z^{P-}_h \p^-_P  \p^-_N \L  \\[5pt]
    \dpp\f_A^M  &=&  \f_A^P \e^{MN} \p^-_P  \p^-_N \L        .\label{eqs}
\end{array}\end{equation}
Once the explicit solutions are known,
\begin{equation} \begin{array}{rll}
	 z_h^{M\pm} &=& z_h^{M\pm}(z_c^{M\pm}, u^\pm) \\[5pt]
	       \f &=& \f(z_c^{M\pm}, u^\pm) ,\end{array}\end{equation}
where the `central' coordinates $z_c^{M\pm}= z^{M\da}u^\pm_\da$ are boundary
values of $z_h^{M\pm}$ satisfying
$\dpp z_c^{M+} =0,\quad \dpp z_c^{M-} = z_c^{M+},$ the required inversion
is straightforward in terms of the matrix of partial derivatives
\begin{equation}  Z^M_N := \pp{z_c^{M-}}{z_h^{N-}}\  .\end{equation}
The explicit supervierbein is constructed by multiplying this by $\f$ on
the left, for the product $\f \cdot Z$ is by construction proportional
to $u^{+\da}u^-_{\db}$, with coefficient being precisely the required
supervierbein independent of the $u$'s :
\begin{equation}     \f^N_A Z^M_N =  u^{+\da}u^-_{\db} E^{M\db}_{A\da}\   .
\end{equation}

Equations \r{eqs} are highly non-linear equations. They however do allow
explicit solution for large classes of $\L$. An explicit illustration of
this procedure may be found in \cite{sdsg,sdg}, where some simple examples
are constructed. The well-known euclidean Taub-NUT solution, for
instance, corresponds to the prepotential $ (x_h^{1+}x_h^{2+} )^2$.
Further examples, including the Atiyah-Hitchin metric, which corresponds to
$\L = (x_h^{1+} )^4 +  (x_h^{2+} )^4$, are currently under investigation in
collaboration with S. Shnider and J. Schiff.

Now, the explicit solubility of the equations \r{ssd} raises an important
question. For just as harmonic space is an auxiliary space in which the
entire content of these theories is encoded in free holomorphic data and
we need to have the above decoding procedure to obtain the supervierbein,
superspace in turn is also basically an auxiliary space in which
supercovariance is manifest. The equations \r{ssd} actually describe not only
a graviton with self-dual Riemann tensor, but also a gravitino satisfying the
Rarita--Schwinger equation, a Yang-Mills field satisfying the Yang-Mills
self-duality equation, a spin
half field satisfying the Weyl equation and scalar fields satisfying the
covariant d'Alembert equation (and for higher N there are further fields
satisfying appropriate first-order equations). All these fields are not
free, but are coupled rather non-trivially with each other. That all this
fits into the neat--looking set of super curvature constraints \r{ssd}
is rather remarkable. However, it also means that the mapping from the set
of component fields satisfying these various field equations to the
supervierbein satisfying \r{ssd} is rather nontrivial and we need to ensure
that it is a 1--1 mapping. What this means is that we must be able to
systematically unravel a unique set of component solutions in $x$-space from
any explicit supervierbein in superspace satisfying \r{ssd}; and conversely,
given any set of fields satisfying the component equations, there must exist
a unique supervierbein satisfying the constraints \r{ssd}. A very efficient
procedure for thus decoding the component content of supervierbeins uses
a generalisation of a technique developed for super Yang-Mills theories
in \cite{hhls}.
The crucial idea is an appropriate choice of gauge. When we have a
description in terms of superfields, supercovariance is manifest since the
local parameters of diffeomorphisms and tangent superrotations depend on
superspace coordinates $x$ and $\bt$. Now, in order to perform component
expansions, we need to choose a gauge which allows us to treat the $\bt$'s
as expansion parameters rather than independent variables. A gauge which does
this, which eliminates all $\bt$-dependence of diffeomorphism and
gauge transformation parameters, without however tampering with the standard
$x$-dependent invariances of the component theory, is the condition that the
Euler operator which measures the degree of homogeneity in the odd variables
is equal to its covariant version, i.e.
\begin{equation}
\D \equiv \bt^{m\db} \der{\bt^{m\db}} = \bt^{m\db} \n_{m\db}
      = \bt^{m\db} E^{A\dg}_{m\db} \n_{A\dg}\   .\end{equation}
This is tantamount to the following conditions for the components of the
supervierbein and superconnection:
\begin{equation}\begin{array}{rll}  \bt^{m\db} E^{g\dg}_{m\db} &=& 0 \\[5pt]
\bt^{m\db} \o_{m\db}  &=& 0 \\[5pt]
E^{k\dg}_{m\db}  &=& \d^k_m\d^\dg_\db + E^{\g\da}_{m\db} F^{k\dg}_{\g\da}\
        ,\end{array}\end{equation}
for some $F^{k\dg}_{\g\da}$. In this gauge the constraints yield the
following action of the Euler operator $\D$ on the supervierbein components:
\begin{equation}\begin{array}{rll}
\D E^{B\dg}_{\nu\db}  &=& \e^{BA} \bt^{k\dg} (\o_{\nu\db} )_{kA}\\[5pt]
(1+\D) E^{B\dg}_{n\db} - \d^B_n \d^\dg_\db &=&
                     - \e^{BA} \bt^{k\dg} (\o_{n\db} )_{kA}\\[8pt]
\D (\o_{\nu\db} )_{AB} &=&   \bt^m_\db R_{m\nu AB} \\[5pt]
(1+\D) (\o_{n\db} )_{AB} &=&   \bt^m_\db R_{mnAB}\  .
\label{D}\end{array}\end{equation}
Now, since the right-hand sides explicitly contain the odd coordinate
linearly, in a $\bt$-expansion the $k$-th order terms on the left-hand sides
are given by the $(k-1)$-th order terms of the superfield expressions
multiplying $\bt$
on the right-hand sides. These relations therefore recursively define the
superfields from their leading components; and they do so in a unique manner.
By repeated application of these relations, the leading components may
be seen to determine the entire $\bt$-expansion of the superfields.
The leading
components of some of the supercurvatures on the right-hand side of \r{ssd}
may thus be seen to be the fields themselves rather than field-strengths.
For instance, the dimension $-\half$ supercurvature leads with the spin
$\half$ field,
\begin{equation} R_{m\nu ni} = \l_{\nu[mni]} + \dots  \end{equation}
and the dimension $0$ curvature contains a string of component fields of
higher and higher spin (depending on the value of the extension N of
superspace) leading with the scalar field,
\begin{eqnarray}  R_{mnil}& =& \f_{mnil}  +\bt^{p\da} \chi_{\da pmnil}
\nonumber\\[5pt] &&  + \half  \bt^{p\da}\bt^{q\db} g_{\da\db pqmnil}
\nonumber\\[5pt] &&
 + {1\over 6}\bt^{p\da}\bt^{q\db}\bt^{r\dg} h_{\da\db\dg qprmnil}
\nonumber\\[5pt]
 &&+ {1\over 24}\bt^{p\da}\bt^{q\db}\bt^{r\dg}\bt^{s\dd}
      C_{\da\db\dg\dd pqrsmnil} + \dots \nonumber\end{eqnarray}
The supervierbein therefore leads, in a
$\bt$-expansion  thus:
\begin{eqnarray}  E^{A\da}_{N\db} &=& \nonumber\end{eqnarray} $$ \pmatrix{
e^{\a\da}_{\nu\db} + \dots  &
\psi^{i\da}_{\nu\db} + \eta^{ik}\bt^{m\da} A_{\nu\db km} + .. \cr
\bt^{i\da}\bt^m_\db \l^\a_{mni} + .. &
\d^i_n \d^\da_\db -  \bt^{j\da}\bt^{m\db}\eta^{ik} \f_{ijkn} + .. \cr} $$
where $e$ is the self-dual vierbein (graviton), $\psi$ is the self-dual spin
${3\over 2}$ gravitino, $A$ is the self-dual vector potential, $\l$ is a
spin $\half$ field and $\f$ is the scalar field. All further fields in the
supermultiplet occur at higher orders in the lower right-hand corner; and
they all have a unique position in this $\bt$-expansion. In fact, as for
self-dual Yang-Mills theories \cite{uncon}, consistency does not seem to
require any limit on the value of the extension N and \r{ssd} thus yields
consistent component equations for fields of arbitrarily high spin.
A detailed analysis is currently under way and will be reported on shortly.

\goodbreak
\end{document}